\documentclass[twocolumn,aps,prb,showpacs,superscriptaddress,preprintnumbers,amsmath,amssymb]{revtex4}
\usepackage[latin9]{inputenc}
\usepackage{float}
\usepackage{bm}
\usepackage{amsmath}
\usepackage{amssymb}
\usepackage{graphicx}
\usepackage{epstopdf}
\makeatletter

\providecommand{\tabularnewline}{\\}

\@ifundefined{textcolor}{}
{%
 \definecolor{BLACK}{gray}{0}
 \definecolor{WHITE}{gray}{1}
 \definecolor{RED}{rgb}{1,0,0}
 \definecolor{GREEN}{rgb}{0,1,0}
 \definecolor{BLUE}{rgb}{0,0,1}
 \definecolor{CYAN}{cmyk}{1,0,0,0}
 \definecolor{MAGENTA}{cmyk}{0,1,0,0}
 \definecolor{YELLOW}{cmyk}{0,0,1,0}
}

\usepackage{dcolumn}
\usepackage{bm}

\makeatother

\begin{document}

\title{Robustness of Quantum Spin Hall Effect in an External Magnetic Field}

\author{Song-Bo Zhang}

\affiliation{Department of Physics, The University of Hong Kong, Pokfulam Road,
Hong Kong, China}

\author{Yan-Yang Zhang}

\affiliation{SKLSM, Institute of Semiconductors, Chinese Academy of Sciences,
P.O. Box 912, Beijing 100083, China}

\affiliation{Department of Physics, The University of Hong Kong, Pokfulam Road,
Hong Kong, China}

\author{Shun-Qing Shen}

\affiliation{Department of Physics, The University of Hong Kong, Pokfulam Road,
Hong Kong, China}
\begin{abstract}
The edge states in the quantum spin Hall effect are expected to be
protected by time reversal symmetry. The experimental observation
of the quantized conductance was reported in the InAs/GaSb quantum
well {[}Du et al, arXiv:1306.1925{]}, up to a large magnetic field,
which raises a question on the robustness of the edge states in the
quantum spin Hall effect under time reversal symmetry breaking. Here
we present a theoretical calculation on topological invariants for
the Benevig-Hughes-Zhang model in an external magnetic field, and
find that the quantum spin Hall effect retains robust up to a large
magnetic field. The critical value of the magnetic field breaking
the quantum spin Hall effect is dominantly determined by the band
gap at the $\Gamma$ point instead of the indirect band gap between
the conduction and valence bands. This illustrates that the quantum
spin Hall effect could persist even under time reversal symmetry breaking. 
\end{abstract}


\maketitle

\section{Introduction}

The quantum spin Hall effect (QSHE) is a novel state of quantum matter,
in which an electric field can generate a transverse spin current
\cite{Moore-10nature,Hazan-10rmp,Qi-11rmp,Shen-book}. A quantum spin
Hall system has a bulk gap between the conduction and valence bands
meanwhile processing a pair of gapless helical edge states surrounding
the boundaries\cite{Kane-05prla,Bernevig-06science,Liu-08prl}. The
gapless helical edge states give rise to a quantized conductance in
a two-terminal measurement, which has been observed experimentally
in HgTe/CdTe quantum well \cite{Konig-07science} and in InAs/GaSb
quantum well \cite{Knez-11prl}. While these edge states are expected
to be protected by time reversal symmetry \cite{Kane-05prlb}, a recent
measurement of QSHE in the InAs/GaSb quantum well surprisingly indicates
that the quantized plateau of conductance persists up to a 12 T (Tesla)
in-plane magnetic field, or an 8 T perpendicular magnetic field\cite{Du-13xxx,Du2014}.
This observation raises a question on the robustness of QSHE under
time reversal symmetry breaking.

The electronic backscattering in the gapless edge states is prohibited
by time reversal symmetry, so that the transport is robust against
disorders respecting the symmetry\cite{Xu-06prb,Wu-06prl,Onoda-07prl}.
An external magnetic field breaks time reversal symmetry and leads
to two important effects: a Peierls phase to the orbital motion, and
a Zeeman split to the spin motion. The interplay between the spin-orbit
coupling and the Zeeman coupling may break QSHE, where the edge states
open a small sub-gap \cite{Konig-08JPSJ,Yang-11prl}, and the robust
transports for QSHE break down similar to the result caused by the
finite size effect\cite{Zhou-08prl}.

In this paper, we present theoretical calculations on topological
invariants of the Bernevig-Hughes-Zhang (BHZ) model for the QSHE,
under an external magnetic field. Our main work focuses on the orbital
motion (i.e., the Landau level forming) effects of magnetic field,
ignoring the Zeeman couplings. This is reasonable for QSHE materials
with small $g$ factors. Furthermore, due to the absence of spin-flip
(e.g., Rashba like spin-orbital coupling) terms, the physical spin
$S_{z}$ is preserved and the system can be decoupled into two components
with opposite $S_{z}$. This makes the two spin-dependent Chern numbers
(for spin up and down components respectively) well defined, even
in the presence of an external field. Correspondingly, it is found
that the edge states persist up to a large magnetic field, until some
band crossing happens. The band gap at the $\Gamma$ point instead
of the indirect gap plays an important role in determining the critical
magnetic field to break down the QSHE. However, when the symmetry
preserving $S_{z}$ is also broken by spin-flip terms coupling two
spin components, usually the edge states are no longer robust. At
the end the effect of a finite Zeeman coupling for perpendicular field is studied, and the critical values the Zeeman
field are presented. The effects of other spin-orbital coupling and the in-plane Zeeman field are also discussed.

\section{Model and solutions}

We start with the BHZ model defined, in the basis $\{\left|s\uparrow\right\rangle ,\left|p\uparrow\right\rangle ,\left|s\downarrow\right\rangle ,\left|p\downarrow\right\rangle \}$,
for the QSHE in quantum wells \cite{Bernevig-06science}, 
\begin{equation}
H_{0}(\mathbf{k})=\left(\begin{matrix}h_{+}(\mathbf{k}) & 0\\
0 & h_{-}(\mathbf{k})
\end{matrix}\right)\label{eq:model}
\end{equation}
where $h_{\pm}(\mathbf{k})=\epsilon(\mathbf{k})\sigma_{0}+\bm{d}_{\pm}\cdot\bm{\sigma}$
with denoting $\mathbf{d}_{\pm}=[\pm Ak_{x},-Ak_{y},\mathcal{M}(\mathbf{k})]$,
$\epsilon(\mathbf{k)}=-Dk^{2}$, $\mathcal{M}(\mathbf{k})=\Delta-Bk^{2}$,
$k^{2}=k_{x}^{2}+k_{y}^{2}$, $k_{\pm}=k_{x}\pm ik_{y}$, and $\mathbf{\sigma}_{i}$
are the Pauli matrices for the orbital $\{\left|s\right\rangle ,\left|p\right\rangle \}$.
The system possesses time reversal symmetry implied by the relation
between two spin components, $h_{-}(\mathbf{k})=h_{+}^{*}(-\mathbf{k})$.
This model has been used to describe the QSHE in HgTe/CdTe and InAs/GaSb
quantum wells \cite{Bernevig-06science,Liu-08prl}.

The Hamiltonian (\ref{eq:model}) can be exactly diagonalized, and
two branches of doubly degenerated eigenenergies are: 
\begin{equation}
E_{s}^{\pm}(\mathbf{k})=\epsilon(\mathbf{k})+s|\mathbf{d}_{\pm}|=\epsilon(\mathbf{k})+s\sqrt{M^{2}(\mathbf{k})+A^{2}k^{2}},
\end{equation}
where $s=+1$ ($-1$) stands for the conduction (valence) band, and
the superscript $\pm$ of $E_{s}^{\pm}(\bm{k})$ stands for spin up
(down) component,which are degenerated here. The term $\epsilon(\bm{k})$
breaks the particle-hole symmetry. When $|D|^{2}<|B|^{2}$, there
is an energy gap between the conduction band $E_{s=+1}^{\pm}(\bm{k})$
and the valence band $E_{s=-1}^{\pm}(\bm{k})$. The gap at $\Gamma$
($\bm{k}=0$) is determined by $2\Delta$. According to the $Z_{2}$
classification, the system can be classified as a topologically trivial
insulator ($B\cdot\Delta<0$) or nontrivial ($B\cdot\Delta>0$) insulator\cite{Lu-10prb}
when the Fermi level is located within the gap. For a large value
of $\left|\Delta\right|$, the band gap (the minimum band separation
in the Brillouin zone, which may not be located at $\Gamma$) is approximately
given by $2|A|\sqrt{\Delta/B}$ when $B\cdot\Delta>0$ as shown in
Fig. 1.

Due to the decoupling between the two spin components $h_{\pm}$,
the physical spin operator $\mathbf{S}_{z}=\tau_{z}\otimes\sigma_{0}$
commutes with $H_{0}$, and has good quantum numbers $S_{z}=\pm1$,
where $\tau_{i}$ refer to Pauli matrices for physical spin and $\sigma_{0}$
is a $2\times2$ identity matrix. This symmetry guarantees that for
each spin component ($h_{+}$ or $h_{-}$), a spin-dependent Chern
number is well defined. The Chern numbers for these two spin components
($\pm$) are 
\begin{equation}
n_{\pm}=\pm\frac{1}{2}[\mathrm{sgn}(B)+\mathrm{sgn}(\Delta)],\label{EqZeroB}
\end{equation}
when the Fermi energy lies in the bulk gap\cite{Lu-10prb}. As a consequence,
the Hall conductance for the whole system is always zero, $\sigma_{xy}=\left(n_{+}+n_{-}\right)\frac{e^{2}}{h}=0$
due to time reversal symmetry, as expected. On the other hand, the
spin Chern number \cite{Sheng-06prl} defined as $n_{s}=\left(n_{+}-n_{-}\right)/2$
equals $1$ or $-1$ when $B\cdot\Delta>0$ , which indicates the
spin Hall conductance is $\sigma_{s}=n_{s}\frac{e}{4\pi}$, and the
system exhibits the QSHE.

\begin{figure}[H]
\centering \includegraphics[width=8.5cm]{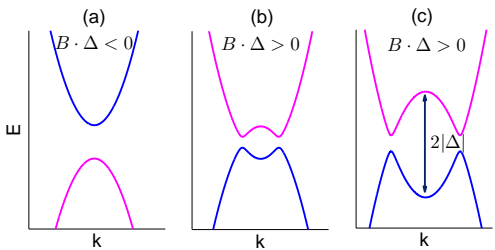} \protect\caption{Schematic of two-dimensional energy spectra of the quantum well systems.
(a) the normal state of $B\cdot\Delta<0$. (b) the band inverted state
of $B\cdot\Delta>0$. (c) the band inverted state of $B\cdot\Delta>0$
with a large gap at $k=0$ but a small bulk gap such as in the InAs/GaSb
quantum well.}
\label{figure1} 
\end{figure}

\begin{table}[!hbp]
\begin{tabular}{|c|c|c|c|c|}
\hline 
Parameters  & $\Delta$ {[}eV{]}  & B {[}eV$\mathrm{nm}^{2}${]}  & D {[}eV$\mathrm{nm}^{2}${]}  & A {[}eV$\mathrm{nm}${]} \tabularnewline
\hline 
HgTe  & -0.010  & -0.686  & -0.512  & 0.365 \tabularnewline
\hline 
InAs/GaSb  & -0.008  & -0.400  & -0.300  & 0.023 \tabularnewline
\hline 
\end{tabular}\protect\caption{The Bernevig-Hughes-Zhang model parameters for quantum wells in the
band inverted regime. Parameters for HgTe/CdTe quantum well are at
$d=7$ nm \cite{Konig-08JPSJ}. For the InAs/GaSb quantum well, $\Delta$
is deduced from the experiment, and other parameters are taken to
have a ratio of the $2\Delta$ to to the indirect gap about 4 \cite{Knez-11prl}. }

\label{parameter_table} 
\end{table}

A perpendicular magnetic field $\mathbf{B}=\mathcal{B}\mathbf{\hat{z}}$
(we assume $\mathcal{B}>0$ without losing generality) breaks the
time reversal symmetry, but $\mathbf{S}_{z}$ is still preserved,
which makes it possible to calculate the Chern numbers for each spin
component $h_{\pm}$, separately. In the absence of the Zeeman splitting,
the magnetic field only manifests itself on the orbital motion. Similar
to the case of two-dimensional electron gas in a perpendicular magnetic
field\cite{Shen-04prl}, the wave vector in Eq. (\ref{eq:model})
is replaced by the substitution, $\mathbf{k}\rightarrow-i\nabla_{\mathbf{r}}+e\mathbf{A}/\hbar$,
where $\mathbf{A}$ is the vector potential so that $\mathbf{B}=\nabla\times\mathbf{A}$.
Choosing the Landau gauge $\mathbf{A}=(-\mathcal{B}y,0,0)$, which
preserves the translational symmetry in $x$ direction, we can take
the eigen wavefunction as the form $\varphi(x,y)=e^{ikx}\phi(y)$.
We define the ladder operators in the following form, 
\begin{equation}
a^{\dagger}(k)=\dfrac{1}{\sqrt{2}}(\dfrac{y-y_{0}}{\ell_{B}}-\ell_{B}\partial_{y}),
\end{equation}
\begin{equation}
a(k)=\dfrac{1}{\sqrt{2}}(\dfrac{y-y_{0}}{\ell_{B}}+\ell_{B}\partial_{y}),
\end{equation}
where $y_{0}=\ell_{B}^{2}k$ is the guiding center of the wave package,
and $\ell_{B}=\sqrt{\hbar/e\mathcal{B}}$ is the magnetic length.
These two operators obey the canonical commutation relation, $[a(k),a^{\dag}(k^{\prime})]=1$.
Thus, the Hamiltonian can be re-expressed in terms of the ladder operators
as 
\begin{equation}
\begin{array}{cc}
h_{\pm}(a,a^{\dagger}) & =\omega_{2}(a^{\dagger}a+\dfrac{1}{2})\sigma_{0}\pm\eta(a^{\dagger}\sigma_{+}+a\sigma_{-})\\
 & +\left[\Delta+\omega_{1}(a^{\dagger}a+\dfrac{1}{2})\right]\sigma_{z}
\end{array}\label{eq:6}
\end{equation}
with $\eta=-\sqrt{2}A/\ell_{B}\dfrac{}{}$, $\omega_{1}=-2B/\ell_{B}^{2}$
and $\omega_{2}=-2D/\ell_{B}^{2}$. It is now easy to analytically
solve the eigen-problems for $h_{\pm}$, respectively.

For the spin-up component $h_{+}$, the eigenenergies are 
\begin{equation}
\begin{aligned}E_{n,k,s}^{+}= & \dfrac{1}{2}[\omega_{1}+2n\omega_{2}+s\sqrt{(\omega_{2}+2(\Delta+n\omega_{1}))^{2}+4n\eta^{2}}],\end{aligned}
\label{Landau_levels}
\end{equation}
where $\eta$ and $\omega_{1,2}$ are the functions of $\mathcal{B}$.
The corresponding eigenstates $\left|n,k,s\right\rangle _{+}$ have
a degeneracy $N_{\phi}=\Omega/(2\pi\ell_{B}^{2})$ with $\Omega$
the area of the two-dimensional system for different wave vectors
$k$. Explicitly, the two-component eigen-function for $h_{+}$ is
given by 
\begin{equation}
\left|n,k,s\right\rangle _{+}=\left(\begin{array}{c}
\cos\theta_{ns}\phi_{n,k}\\
\sin\theta_{ns}\phi_{n-1,k}
\end{array}\right)\label{eigenstates}
\end{equation}
Where $\phi_{n,k}(y)=\frac{1}{\sqrt{n!2^{n}\ell_{B}\sqrt{\pi}}}e^{ikx-(y-y_{0})^{2}/2\ell_{B}^{2}}\mathcal{H}_{n}(\frac{y-y_{0}}{\ell_{B}})$
is the eigenstate of the number operator $a^{\dag}a$ with an integer
eigenvalue $n$, and $\mathcal{H}_{n}$ are the Hermite polynomials
defined as $\mathcal{H}_{n}({\xi})=(-1)^{n}e^{\xi^{2}}\frac{\partial^{n}}{\partial\xi^{n}}e^{-\xi^{2}}$.
Notice there are two Landau levels with $s=\pm$ for $n\geqslant1$,
but only $s=\mathrm{sgn}\left(\omega_{2}+2\Delta\right)$ for $n=0$.
For $n=0$, $\theta_{n=0,s}=0$. Otherwise, for $n\geqslant1$, $\tan\theta_{n,s}=s\sqrt{1+u_{n}^{2}}-u_{n}$
with $u_{n}=\left[\omega_{2}+2(\Delta+n\omega_{1})\right]/\sqrt{4n\eta^{2}}$.

For the spin-down component $h_{-}$, the eigenenergies are 
\begin{equation}
\begin{aligned}E_{n,k,s}^{-}= & \dfrac{1}{2}[2n\omega_{2}-\omega_{1}+s\sqrt{(\omega_{2}-2(\Delta+n\omega_{1}))^{2}+4n\eta^{2}}],\end{aligned}
\label{Landau_levels-1}
\end{equation}
and the corresponding two-component eigen-function for $h_{-}$ is
given by 
\begin{equation}
\left|n,k,s\right\rangle _{-}=\left(\begin{array}{c}
-\sin\theta_{n,s}^{-}\phi_{n-1,k}\\
\cos\theta_{n,s}^{-}\phi_{n,k}
\end{array}\right)\label{eigenstates-1}
\end{equation}
Similarly, there are two Landau levels with $s=\pm$ for $n\geqslant1$,
but only $s=\mathrm{sgn}\left(\omega_{2}-2\Delta\right)$ for $n=0$.
For $n=0$, $\theta_{n=0,s}^{-}=0$. Otherwise, for $n\geqslant1$,
$\tan\theta_{n,s}^{-}=s\sqrt{1+v_{n}^{2}}-v_{n}$ with $v_{n}=\left[\omega_{2}-2(\Delta+n\omega_{1})\right]/\sqrt{4n\eta^{2}}$.

When the Fermi energy lies in the bulk gap, as in the following calculations,
the Landau levels {[}Eq. \eqref{Landau_levels}{]} are divided into
two sets: the electron-like Landau levels $E_{n,k,s=+}^{m}$ evolving
from the conduction band, and the hole-like Landau levels $E_{n,k,s=-}^{m}$
evolving from the valence band for $n>0$, with the superscript $m=\pm$
referring to spin components $h_{\pm}$. The $0$th Landau level of
$h_{+}$ is electron-like (hole-like) if $2\Delta+\omega_{2}>0$ ($<0$),
so we can denote it as $E_{0,k,+}^{+}=\frac{\omega_{1}}{2}+\frac{\omega_{2}}{2}+\Delta$
($E_{0,k,-}^{+}=\frac{\omega_{1}}{2}+\frac{\omega_{2}}{2}+\Delta$).
Analogously, for the $h_{-}$, the $0$th Landau level is electron-like
(hole-like) if $2\Delta-\omega_{2}<0$ ($>0$), and we denote $E_{0,k,+}^{-}=-\frac{\omega_{1}}{2}+\frac{\omega_{2}}{2}-\Delta$
($E_{0,k,-}^{-}=-\frac{\omega_{1}}{2}+\frac{\omega_{2}}{2}-\Delta$).
There is always a finite gap between the electron-like Landau levels
and the hole-like Landau levels. In other words, the gap between the
conduction and valence bands never close in a finite magnetic field
when the Zeeman splitting is ignored.

The developments of the Landau levels $E_{n,k,s}^{+}(\mathcal{B})$
and $E_{n,k,s}^{-}(\mathcal{B})$ under the magnetic field for some
typical parameters associated with realistic quantum wells are plotted
in Fig.~2. One important observation in Figs.~2(a), (b), (d) and
(e) is that, for one spin component, the Landau levels from the conduction
band (electron-like) $E_{n,k,s}^{\pm}(\mathcal{B})$ never cross with
those from the valance band (hole-like) $E_{n,k,-s}^{\pm}(\mathcal{B})$.
In other words, the orbital motion of electrons in a magnetic field
cannot lead to the gap closing between the conduction and valence
bands. This is one of the main results in this work. As a result,
with the increasing of magnetic field $\mathcal{B}$, possible topological
transitions (band crossings) in the bulk gap can happen only when
both spin components are considered together, as shown in Fig.~2(c)
and (f). The calculation of the spin Chern numbers is presented in
the following section.

\begin{widetext}

\begin{figure}[H]
\centering \includegraphics[width=15cm]{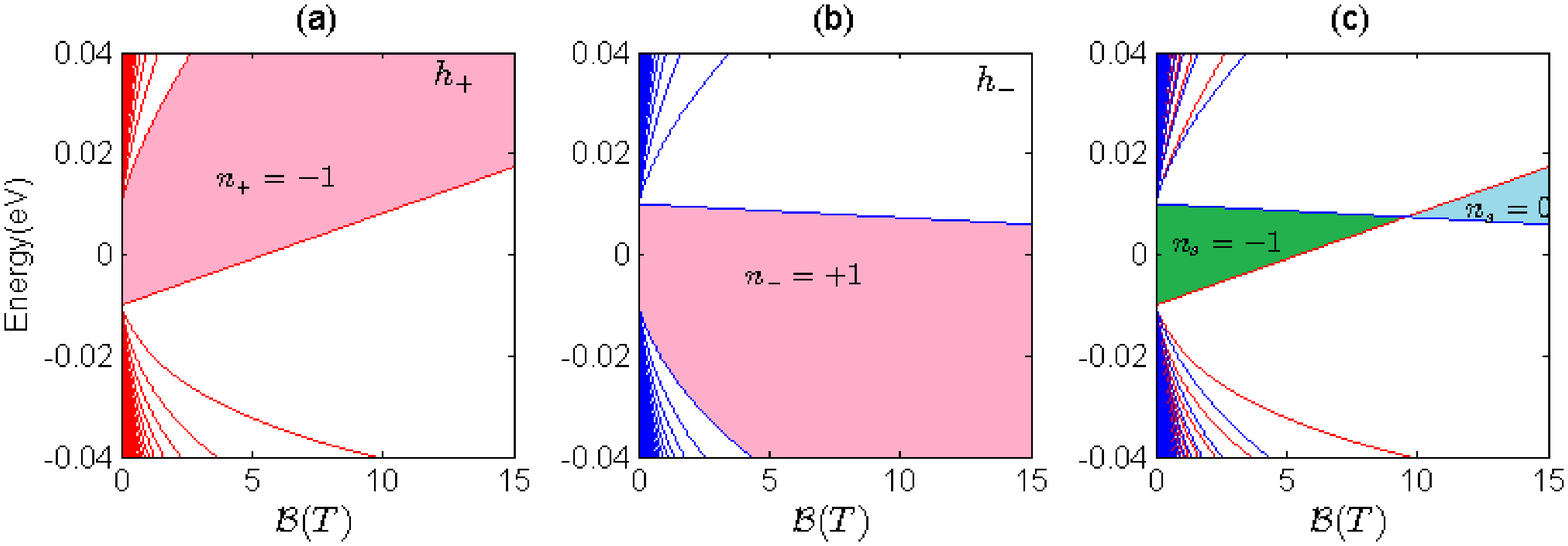} \includegraphics[width=15cm]{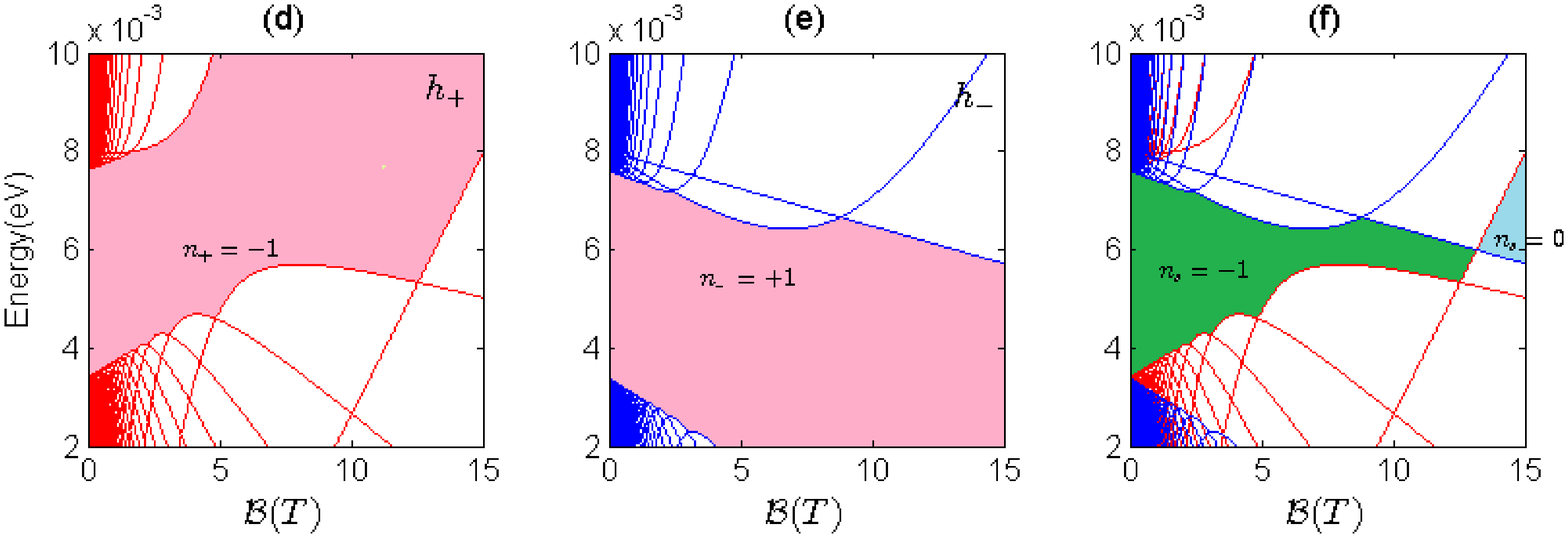}
\protect\caption{The Landau levels as functions of a perpendicular magnetic field $\mathcal{B}$
in a unit of tesla (T). The upper row is for the HgTe quantum well
and the lower row is for the InAs/GaSb quantum well. The left, middle
and right columns correspond to $h_{+}$ (spin up), $h_{-}$ (spin
down) and $H_{0}$ (whole system), respectively. The shaded region
is the bulk gap between the conduction and valence band, with topological
invariants labelled. Model parameters are listed in Table.\ref{parameter_table}.}
\label{figure2} 
\end{figure}

\end{widetext}

\section{Chern numbers in a magnetic field}

The main result in this section is summarized as follows: in the presence
of a perpendicular magnetic field, when the Fermi level lies in the
bulk gap of $h_{+}$ and $h_{-}$, the spin-dependent Hall conductivities
are always 
\begin{equation}
\sigma_{xy}^{\pm}\equiv n_{\pm}\frac{e^{2}}{h}=\pm\frac{1}{2}\left[\mathrm{sgn}(B)+\mathrm{sgn}(\Delta)\right]\frac{e^{2}}{h}.\label{General_Chern_Number}
\end{equation}
It means that for any finite field $\mathcal{B}$, $\sigma_{\pm}=0$
if $B\cdot\Delta<0$ while $\sigma_{\pm}=\mp e^{2}/h$ if $B\cdot\Delta>0$
when the Fermi level is located within the band gap between the conduction
and valence bands. These results, which can be calculated explicitly
from the Kubo formula, or the formula for Chern number \cite{Lu-10prb},
are identical to those in the absence of magnetic field, i.e., $\mathcal{B}=0$.

\subsection{A general expression}

In the following, we present the detailed calculations of these Hall
conductivities at zero temperature. By using the Kubo formula, the
Chern numbers for spin component $m=\pm$ are given by 
\begin{equation}
\begin{aligned}n_{m}= & -\dfrac{2\pi\hbar^{2}}{\Omega}\sum_{k}\sum_{n\neq n^{\prime},s,s'}\left[f(E_{n,k,s}^{m}-\mu)-f(E_{n',k',s'}^{m}-\mu)\right]\\
 & \dfrac{\mathrm{Im}\left[_{m}\left\langle n,k,s\right|\hat{v}_{x}^{m}\left|n',k,s'\right\rangle _{mm}\left\langle n',k,s'\right|\hat{v}_{y}^{m}\left|n,k,s\right\rangle _{m}\right]}{(E_{n,k,s}^{m}-E_{n',k,s'}^{m})^{2}},
\end{aligned}
\label{Kubo_formula}
\end{equation}
where $f(x)$ is the Fermi-Dirac distribution function, $\mu$ is
the Fermi energy and the velocity operators can be obtained by evaluating
$\hat{v}_{x}^{\pm}=\partial h_{\pm}/\hbar\partial k_{x}$ and $\hat{v}_{y}^{\pm}=\partial h_{\pm}/\hbar\partial k_{y}$
with the substitution $\mathbf{k}\rightarrow-i\nabla_{\mathbf{r}}+e\mathbf{A}/\hbar$,
\begin{eqnarray}
\hat{v}_{x}^{\pm}=-\dfrac{\ell_{B}}{\sqrt{2}\hbar}\left(\begin{matrix}\omega_{+}(a^{\dag}+a) & \pm\eta\\
\pm\eta & \omega_{-}(a^{\dag}+a)
\end{matrix}\right),\label{velocity_x1}\\
\hat{v}_{y}^{\pm}=\dfrac{i\ell_{B}}{\sqrt{2}\hbar}\left(\begin{matrix}\omega_{+}(a^{\dag}-a) & -\eta\\
+\eta & \omega_{-}(a^{\dag}-a)
\end{matrix}\right),\label{velocity_y1}
\end{eqnarray}
where $\omega_{\pm}=\pm\omega_{1}+\omega_{2}$.

Assume that the Fermi energy be located within the gap between the
electron-like and the hole-like Landau levels. Then $f(E_{n,k,s}^{m}-\mu)=0$
if $E_{n,k,s}^{m}$ is electron-like ($s=+$), while $f(E_{n,k,s}^{m}-\mu)=1$
if $E_{n,k,s}^{m}$ is hole-like ($s=-$). The summation over $k$
in Eq.(\ref{Kubo_formula}) only gives a factor of the Landau degeneracy
$N_{\Phi}=\Omega/(2\pi\ell_{B}^{2})$ for each Landau level. Thus
the Chern number eventually becomes 
\begin{equation}
\begin{aligned}n_{m}= & \dfrac{\hbar^{2}}{\ell_{B}^{2}}\sum_{n,n^{\prime}=0}^{\infty}\left\{ \mathrm{Im}[v_{x;n-,n^{\prime}+}^{m}v_{y;n^{\prime}+,n-}^{m}]\dfrac{1}{(E_{n,k,-}^{m}-E_{n^{\prime},k,+}^{m})^{2}}\right.\\
 & -\left.\mathrm{Im}[v_{x;n+,n^{\prime}-}^{m}v_{y;n^{\prime}-,n+}^{m}]\dfrac{1}{(E_{n,k,+}^{m}-E_{n^{\prime},k,-}^{m})^{2}}\right\} \\
= & \dfrac{2\hbar^{2}}{\ell_{B}^{2}}\sum_{n,n^{\prime}=0}^{\infty}\mathrm{Im}[v_{x;n-,n^{\prime}+}^{m}v_{y;n^{\prime}+,n-}^{m}]\dfrac{1}{(E_{n,k,-}^{m}-E_{n^{\prime},k,+}^{m})^{2}},
\end{aligned}
\label{Kubo_formula_simplify}
\end{equation}
where we have denoted the velocity matrix elements as $v_{i;ns,n^{\prime}s^{\prime}}^{m}\equiv~_{m}\langle n,k,s|\hat{v}_{i}^{m}|n^{\prime},k,s^{\prime}\rangle_{m}$
with $i=x,y$. These velocity matrix elements are evaluated with the
help of the expressions in Eq. (\ref{Landau_levels}) for the Landau
levels as well as Eqs. (\ref{velocity_x1}) and (\ref{velocity_y1}).
It turns out that $v_{i;ns,n^{\prime}s^{\prime}}^{m}$ are nonzero
only if $n^{\prime}=n\pm1$. By evaluating the matrix elements, the
general expressions for the Chern numbers of $h_{\pm}$ are

\begin{widetext} \begin{subequations} 
\begin{align}
n_{+}= & \sum_{n=0}^{\infty}\left\{ \left[\omega_{+}\cos\theta_{n,+}\sin\theta_{n+1,+}\sqrt{n+1}-\omega_{-}\sin\theta_{n,+}\cos\theta_{n+1,+}\sqrt{n}-\eta\cos\theta_{n,+}\cos\theta_{n+1,+}\right]^{2}\dfrac{1}{(E_{n+1,k,-}^{+}-E_{n,k,+}^{+})^{2}}\right.\notag\\
 & \left.-\left[\omega_{+}\cos\theta_{n+1,+}\sin\theta_{n,+}\sqrt{n+1}-\omega_{-}\sin\theta_{n+1,+}\cos\theta_{n,+}\sqrt{n}+\eta\sin\theta_{n+1,+}\sin\theta_{n,+}\right]^{2}\dfrac{1}{(E_{n+1,k,+}^{+}-E_{n,k,-}^{+})^{2}}\right\} ,\label{eq:14a}\\
n_{-}= & \sum_{n=0}^{\infty}\left\{ \left[\omega_{+}\sin\theta_{n,-}^{-}\cos\theta_{n+1,-}^{-}\sqrt{n}-\omega_{-}\cos\theta_{n,-}^{-}\sin\theta_{n+1,-}^{-}\sqrt{n+1}+\eta\cos\theta_{n,-}^{-}\cos\theta_{n+1,-}^{-}\right]^{2}\dfrac{1}{(E_{n+1,k,-}^{-}-E_{n,k,+}^{-})^{2}}\right.\notag\\
 & \left.-\left[\omega_{+}\sin\theta_{n+1,-}^{-}\cos\theta_{n,-}^{-}\sqrt{n}-\omega_{-}\cos\theta_{n+1,-}^{-}\sin\theta_{n,-}^{-}\sqrt{n+1}-\eta\sin\theta_{n+1,-}^{-}\sin\theta_{n,-}^{-}\right]^{2}\dfrac{1}{(E_{n+1,k,+}^{-}-E_{n,k,-}^{-})^{2}}\right\} .\label{eq:14b}
\end{align}
\end{subequations} \end{widetext} In the above summations, special
attention should be paid to the term with $n=0$. For $n_{+}$ if
$2\Delta+\omega_{2}<0$ ($>0$), the $0$th Landau level of $h_{+}$
is hole-like (electron-like). As for $n_{-}$, if $2\Delta-\omega_{2}<0$
($>0$), the $0$th Landau level of $h_{-}$ is electron-like (hole-like)
.

In the presence of the magnetic field, the general analytic expressions
in Eqs.~\eqref{eq:14a} and \eqref{eq:14b} are too long to be simplified
explicitly. Nevertheless we can still draw the conclusion shown in
Eq. (\ref{General_Chern_Number}) in two alternative ways.

\subsection{The case of the particle-hole symmetry}

When the system possesses the particle-hole symmetry, i.e., $D=0$,
the Chern numbers $n_{\pm}$ in Eqs.~\eqref{eq:14a} and \eqref{eq:14b}
can be further reduced to 
\begin{equation}
\begin{aligned}n_{\pm}= & \pm\sum_{n=0}^{\infty}\frac{\omega_{1}(\epsilon_{n+1}-\epsilon_{n})^{2}+2\eta^{2}\Delta-\omega_{1}^{3}-(2n+1)\omega_{1}\eta^{2}}{4\epsilon_{n}\epsilon_{n+1}(\epsilon_{n}+\epsilon_{n+1})},\end{aligned}
\label{Reduced_Hall1}
\end{equation}
where $\epsilon_{n}=\sqrt{(\Delta+n\omega_{1})^{2}+n\eta^{2}}$. By
using the identity $\epsilon_{n+1}^{2}-\epsilon_{n}^{2}=\eta^{2}+\omega_{1}[2\Delta+(2n+1)\omega]$,
the Chern numbers can be rewritten as 
\begin{equation}
\begin{aligned}n_{\pm}= & \pm\dfrac{1}{2}\sum_{n=0}^{\infty}\Big[\dfrac{\Delta+n\omega_{1}}{\epsilon_{n}}-\dfrac{\Delta+(n+1)\omega_{1}}{\epsilon_{n+1}}\Big]\\
= & \pm\dfrac{1}{2}\Big[\dfrac{\Delta}{\epsilon_{0}}-\lim_{n\rightarrow\infty}\dfrac{\Delta+(n+1)\omega_{1}}{\epsilon_{n+1}}\Big]\\
= & \pm\dfrac{1}{2}[\mathrm{sgn}(\Delta)+\mathrm{sgn}(B)].
\end{aligned}
\label{Reduced_Hall2}
\end{equation}
In the last line, we have used 
\begin{equation}
\dfrac{\Delta}{\epsilon_{0}}=\dfrac{\Delta}{\sqrt{\Delta^{2}}}=\mathrm{sgn}(\Delta),
\end{equation}
and 
\begin{equation}
\begin{aligned} & \lim_{n\rightarrow\infty}\dfrac{\Delta+(n+1)\omega_{1}}{\epsilon_{n+1}}=\mathrm{sgn}(\omega_{1})=-\mathrm{sgn}(B).\end{aligned}
\end{equation}
Thus in the case of $D=0$, the Chern numbers can be calculated exactly.

\subsection{The case of the particle-hole symmetry breaking}

When $D\neq0$, so far we could not simplify the expressions explicitly
to the final form as in Eq. (\ref{General_Chern_Number}). However
we can still draw the conclusion based on the following arguments.
First in a weak field limit of $\mathcal{B}$, $|\omega_{1,2}|$,
$|\eta|\ll|\Delta|$, thus 
\begin{equation}
(E_{n+1,k,s}-E_{n+1,k,-s})^{2}\simeq(\epsilon_{n}+\epsilon_{n+1})^{2},\label{approximation1}
\end{equation}
and 
\begin{eqnarray}
\cos\theta_{n,+}\simeq\sin\theta_{n,-}^{-}\simeq\sqrt{\dfrac{1}{2}\Big[1+\dfrac{\Delta+n\omega_{1}}{\epsilon_{n}}\Big]},\label{approximation2}\\
\sin\theta_{n,+}\simeq\cos\theta_{n,-}^{-}\simeq\sqrt{\dfrac{1}{2}\Big[1-\dfrac{\Delta+n\omega_{1}}{\epsilon_{n}}\Big]}.\label{approximation3}
\end{eqnarray}
Substituting Eqs.~\eqref{approximation1}, \eqref{approximation2}
and \eqref{approximation3} into Eqs.~\eqref{eq:14a} and \eqref{eq:14b},
the Chern numbers read \begin{widetext} 
\begin{equation}
\begin{aligned}n_{\pm}= & \dfrac{1}{2}\sum_{n=0}^{\infty}\left\{ \dfrac{1}{\epsilon_{n}}\Big[\pm(\Delta+n\omega_{1})+\dfrac{\omega_{2}(\Delta+n\omega_{1})(2(2n+1)\omega_{1}\pm\omega_{2})}{(\epsilon_{n}+\epsilon_{n+1})^{2}}+\dfrac{n\omega_{2}\eta^{2}}{(\epsilon_{n}+\epsilon_{n+1})^{2}}\Big]\right.\\
 & -\left.\dfrac{1}{\epsilon_{n+1}}\Big[\pm(\Delta+(n+1)\omega_{1})+\dfrac{\omega_{2}(\Delta+(n+1)\omega_{1})(2(2n+1)\omega_{1}\pm\omega_{2})}{(\epsilon_{n}+\epsilon_{n+1})^{2}}+\dfrac{(n+1)\omega_{2}\eta^{2}}{(\epsilon_{n}+\epsilon_{n+1})^{2}}\Big]\right\} .
\end{aligned}
\label{weak_field_approximate}
\end{equation}
\end{widetext} In the weak field limit, i.e., the Landau degeneracy
is 1, 
\begin{equation}
\begin{array}{c}
\left|\dfrac{n\omega_{2}\eta^{2}}{(\epsilon_{n}+\epsilon_{n+1})^{2}}\right|\ll|\Delta+n\omega_{1}|,\\
\left|\dfrac{\omega_{2}(\Delta+n\omega_{1})(\omega_{2}+2(2n+1)\omega_{1})}{(\epsilon_{n}+\epsilon_{n+1})^{2}}\right|\ll|\Delta+n\omega_{1}|,
\end{array}
\end{equation}
thus the Chern numbers in Eq.\eqref{weak_field_approximate} become
\begin{equation}
\begin{aligned}n_{\pm}= & \pm\dfrac{1}{2}\sum_{n=0}^{\infty}\left[\dfrac{\Delta+n\omega_{1}}{\epsilon_{n}}-\dfrac{\Delta+(n+1)\omega_{1}}{\epsilon_{n+1}}\right].\end{aligned}
\end{equation}
Therefore in the weak field limit, the Chern numbers are 
\begin{equation}
\begin{aligned}n_{\pm}= & \pm\dfrac{1}{2}[\mathrm{sgn}(\Delta)+\mathrm{sgn}(B)].\end{aligned}
\end{equation}
These results are identical to the exact ones in the absence of external
magnetic fields, i.e., $\mathcal{B}=0$, as it should be.

Second, for a finite magnetic field, the band gap between the conduction
band and valence band never closes for either $h_{+}$ or $h_{-}$
by increasing the magnetic field. The topological invariant does not
change when there is no band crossing. Therefore the Chern numbers
should remain unchanged in the whole shaded regions in Figs.~2(a),
(b), (d), and (e). The Chern number at a finite magnetic field $\mathcal{B}$
should be equal to that in a weak field limit or $\mathcal{B}=0$
{[}Eq.~(\ref{EqZeroB}){]} , which is just the case for Eq.~\eqref{General_Chern_Number}.
Thus we conclude that the formula holds for $D\neq0$.

Finally we restrict ourselves to one spin component {[}either $h_{+}(\mathcal{B})$
or $h_{-}(\mathcal{B}$){]}. When $B\cdot\Delta>0$ (topologically
nontrivial at $\mathcal{B}=0$), the Chern numbers $n_{\pm}$ in the
bulk gap as shown in the shaded regions in Fig. 2(a), 2(b), 2(d) and
2(e) are either +1 or -1, which is topologically nontrivial. This
is valid for arbitrary value of magnetic field $\mathcal{B}$, as
long as the Fermi energy is located in the bulk gap. The topologically
nontrivial phase in the shaded regions will not be changed without
any band crossing or band inversion.

Alternatively, we can also start with the case of $D=0$ in a finite
field, in which the Chern numbers have been calculated rigorously.
From the dispersion relations in Eq. (2), there always exist a band
gap between the conduction band and valence band even for $D\neq0$.
However, for a large $k$ limit, 
\[
E_{s}^{\pm}(\mathbf{k})=-(D-s\left|B\right|)k^{2}.
\]
When $\left|D\right|\leq\left|B\right|$, there exist an indirect
gap between two bands, but the gap will close when $\left|D\right|>\left|B\right|$.
Thus the Chern numbers are the same when there exist the indirect
band gap. When $\left|D\right|>\left|B\right|$, one band is always
partially filled when the chemical potential $-\left|\Delta\right|<\mu<\left|\Delta\right|$.
Thus the summation in Eq. (\ref{Kubo_formula}) is no longer an integer.
This conclusion can be extended to the case of a finite field.

\subsection{Change of Chern number}

\begin{figure}[!ht]
\centering \includegraphics[width=85mm]{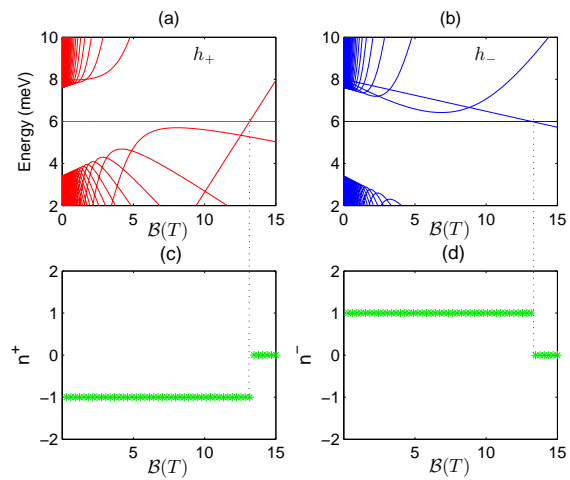} \centering
\caption{ The schematic of evolution of the Chern numbers for each spin component,
with increasing the magnetic field in an InAs/GaSb QW system. The
model parameters are given in the Table. I. The Fermi energy is fixed
at $\mu=6$ meV, as shown by the back lines in (a) and (b). These
Chern numbers (c) and (d) are calculated by using Eqs.~(16a) and
(16b), which is well quantized. The Chern number jumps by +1 ro -1
when the Fermi energy crosses one Landau level. }
\label{figure3} 
\end{figure}

To establish the phase diagram as shown in Fig. 2(c) and (f), we present
the Chern number as a function of the magnetic field $\mathcal{B}$.
While the Chern number is a constant within the gap, it will change
when the Fermi level crosses a Landau level. As a concrete example,
we show what will happen at a definite Fermi energy, with increasing
the magnetic field since this is measurable experimentally. In Fig.~\ref{figure3},
corresponding to Laudau levels for the InAs/GaSb quantum well shown
in Fig.~2(d) and (e), we plot the numerical results for spin Chern
numbers, given that the Fermi energy locates in the bulk gap, i.e.,
$\mu=6$ meV shown by the black lines. Since the increasing of $\mathcal{B}$
does not lead to crossing between the Landau levles originating from
the valance band and those from the conduction band for each spin
component, the spin Chern number does not change, provided that the
Fermi energy locates within the bulk gap. However, the Chern number
changes by 1 or -1 when any Landau level crosses the Fermi energy.
For each spin component, since one Landau level carries a Chern number
$\pm1$, the Chern number of a subsystem will change by $\pm1$ when
one level crosses the Fermi energy. The phase transitions of the whole
system can be determined by counting in both the changes of Chern
numbers for the two spin components, respectively. 

\subsection{Spin Chern numbers and QSHE}

Now let us combine the two spin components of $h_{+}$ and $h_{-}$
as a whole system together. As shown in Fig. 2 (c) and (f), developing
from the identical gap at zero field for both spin components, there
is always a finite region which is just the overlap of their shaded
regions, where $h_{+}$ and $h_{-}$ have opposite and nonzero Chern
numbers. In this region, the total Hall conductance is zero as a summation
of $\sigma_{xy}^{\pm}$, but the spin Hall conductance is still quantized
as $\sigma_{s}=[\left(n_{+}-n_{-}\right)/2]\frac{e}{8\pi}=\left[\mathrm{sgn}(B)+\mathrm{sgn}(\Delta)\right]\frac{e}{8\pi}$
even at a finite magnetic field. In other words, the spin Hall conductance
in the overlapped region in Fig.~\ref{figure2}(c) and (f) remains
as that in the absence of the field. This quantum spin Hall conductance
$\sigma_{s}$ {[}or corresponding spin Chern number $n_{s}=\left(n_{+}-n_{-}\right)/2${]}
is well defined only when there is no coupling between two spin components.

Due to the presence of finite model parameter $B$ (which guarantees
$B\cdot\Delta>0$), the envelops separating the conduction and valence
bands for $h_{+}$ and $h_{-}$ are not parallel to each other, with
the increasing of $\mathcal{B}$. As a result, a topological quantum
phase transition occurs when one Landau level from valence band associated
with one spin component crosses with that from conduction band associated
with another spin component. After this crossing, the Chern numbers
associated with the bulk gap maybe not longer opposite to each other,
$n_{+}\neq-n_{-}$ for two spin components respectively, making the
system from QSHE to other quantum Hall states with both $n=n_{+}+n_{-}\neq0$
and $n_{s}=n_{+}-n_{-}\neq0$

\begin{figure}[ht!]
\centering \includegraphics[width=9cm]{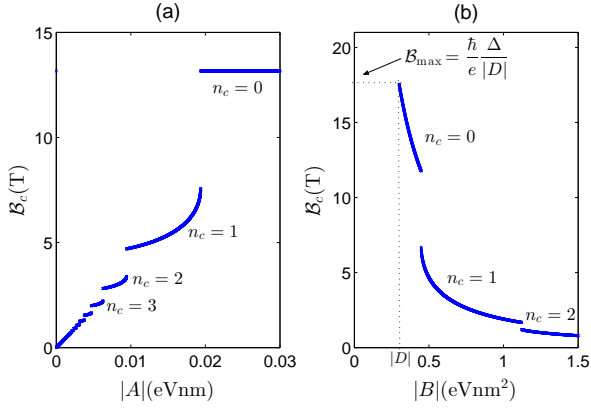} \protect\caption{The first crossing magnetic field $\mathcal{B}_{c}$ as a function
of model parameters for InAs/GaSb quantum well when other parameters
are fixed as in Table.\ref{parameter_table}. (a) $\mathcal{B}_{c}$
as a function of $|A|$. (b) $\mathcal{B}_{c}$ as a function of $|B|$.
Different branches correspond to different Landau levels (as indexed)
that happen to cross first, starting from low field. }
\label{figure4} 
\end{figure}

\section{Critical magnetic field}

We have demonstarted that, the BHZ model (\ref{eq:model}) displays
a well-defined QSHE under a finite magnetic field $\mathcal{B}$,
until a band crossing between one conduction band and one valence
band from $h_{+}$ and $h_{-}$, respectively, happens at a critical
$\mathcal{B}_{c}$. Thus the magnitude of $\mathcal{B}_{c}$ determines
the robustness of the QSHE. Now we are in a position to determine
this critical magnetic field for various model parameters, with the
above knowledge of Landau levels and Chern numbers at finite $\mathcal{B}$.
First of all, consider Landau levels with the index $n=0$. Notice
there is only one Landau level of $n=0$ from each spin component:
the one from $h_{+}$ belongs to the valance band and the one from
$h_{-}$ belongs to the conductance band. They always cross at

\begin{equation}
\mathcal{B}_{0}=\dfrac{\hbar}{e}\dfrac{\Delta}{B}
\end{equation}
therefore a large ratio between $\Delta$ and $B$ always leads to
a large critical value of magnetic field $\mathcal{B}_{0}$. Notice
that the gap at $\Gamma$ point is determined by $2\Delta$, not the
direct gap away from the $\Gamma$ point. For the model parameters
for HgTe and InAs/GaSb quantum wells listed in Table.\ref{parameter_table},
we find that the boundary of the QSHE region is determined by the
two levels of $n=0$, and obtain the critical magnetic field $\mathcal{B}_{c}=\mathcal{B}_{0}=$9.59~T
and 13.16~T respectively. Such a relatively strong critical magnetic
field is consistent with recent experimental observation in InAs/GaSb
quantum well\cite{Du-13xxx,Du2014}.

In a general case, it is also possible for two Landau levels with
index $n>0$ to cross first when increasing magnetic field. The general
procedure to extract $\mathcal{B}_{c}$ is as follows. It is observed
that two definite Landau levels of the sameindex $n>0$ from different
spin components will cross first . The general condition to determine
the band crossing of two Landau levels with index $n>0$ from $h_{-}$
and $h_{+}$ is given by an integer $n_{c}\leq1/\sqrt{w(4-w)}$ with
$w=A^{2}/[B\Delta(1-D^{2}/B^{2})]$. The corresponding magnetic fields
at these crossing points are 
\begin{equation}
\mathcal{B}_{c}=\dfrac{\hbar}{e}\frac{n_{c}}{4n_{c}^{2}-1}\dfrac{2\Delta}{B}\left(\zeta-\sqrt{\zeta^{2}-1+\frac{1}{4n_{c}^{2}}}\right),\label{eq:field}
\end{equation}
where $\zeta=1-w$. Since a pair of Landau levels with smaller $n$
crosses at larger $\mathcal{B}$, the first crossing point $\mathcal{B}_{c}$
by increasing the magnetic field is given by the maximal integer $n_{c}$
from the above inequality.

\begin{figure}[H]
\centering \includegraphics[width=8.5cm]{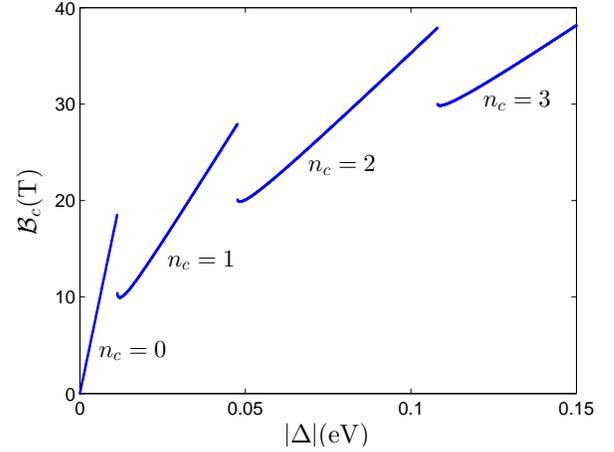} \protect\caption{The critical magnetic field $B_{c}$ as a function of $|\Delta|$.
Different branches correspond to different Landau levels (as indexed)
that happen to cross first.}
\label{figure5} 
\end{figure}

With this process of determining $\mathcal{B}_{c}$, it is natural
to investigate the optimal model parameters for a large critical field,
as a guidance for future device fabrication. For a generic group of
model parameters, the above process is rather tedious, because of
the strongly nonlinear and discontinuous dependence of $\mathcal{B}_{c}$
on these parameters. To extract more insights from Eq.~\eqref{eq:field},
we present the parameter-dependences of the critical magnetic field
in Fig.~\ref{figure4} and Fig.~\ref{figure5}. The critical field
is plotted as a function of the model parameter $A$ in Fig.~\ref{figure4}(a).
For a sufficiently large $A$, the critical field is determined by
the crossing point of two Landau levels of $n_{c}=0$, i.e., $\mathcal{B}_{c}=\dfrac{\hbar}{e}\dfrac{\Delta}{B}$
. When $A$ decreases, $n_{c}$ increases, but the critical field
decreases quickly if other parameters remains unchanged. In Fig.~\ref{figure4}(b),
the critical field increases quickly when the parameter $B$ decreases.
The critical field reaches at a maximal value $\mathcal{B}_{c}=\dfrac{\hbar}{e}\dfrac{\Delta}{\left|D\right|}$
when $\left|B\right|\rightarrow\left|D\right|$. We notice that the
critical field is very sensitive to model parameters $B$ and $D$,
and can be as large as tens of Teslas. As for $\Delta$, as plotted
in Fig.~\ref{figure5}, the first observation is that the critical
field increases with $\Delta$ monotonically for a Landau level crossing
with a specific index $n_{c}$. Increasing $\Delta$ leads to a band
crossing of Landau levels with a higher index, which corresponds to
a larger critical magnetic field $\mathcal{B}_{c}$.

\section{A perpendicular Zeeman field}

The Zeeman coupling can destroy the QSHE in an alternative way. For
a perpendicular Zeeman field, the Hamiltonian has an additional term
\begin{equation}
H_{\bot}=V_{\bot}\left(\begin{array}{cc}
\sigma_{0} & 0\\
0 & -\sigma_{0}
\end{array}\right),
\end{equation}
($\sigma_{0}$ is a $2\times2$ identity matrix) and $S_{z}$ is still
a good quantum number. The main effect of this type of Zeeman coupling
to the BHZ model is the change of the gap in $h_{\pm}$: 
\begin{equation}
\triangle_{\pm}=\triangle\pm V_{\bot}.
\end{equation}
No matter what sign of the gap and the Zeeman field, the Zeeman field
always induces a topological transition to a quantum anomalous Hall
effect at $\left|V_{\bot}^{c}\right|=\left|\triangle\right|$ according
to the formula of the Chern number. When $\left|V_{\bot}^{c}\right|>\left|\triangle\right|$,
one is $0$ while another one is $+1$ or $-1$\cite{Yu-10science,Lu-13prl}.
Thus the total Hall conductance becomes quantized in unit of $e^{2}/h$.

Of course the combined effect of the Zeeman field and the orbital
motion may change the critical values of the magnetic field when the
$g$ factor is not negligible.

\section{Other spin-orbit couplings}

So far we just consider the case that two spin-component $h_{+}$
and $h_{-}$ are decoupled such that the spin Chern numbers are well
defined. When $S_{z}$ is no longer a good quantum number, the problem
becomes more subtle and delicate\cite{Konig-08JPSJ}. For example,
when the Rashba spin-orbit coupling and the Zeeman coupling co-exist
in the Kane-Mele model for QSHE, the helical edge states may open
a tiny sub-gap \cite{Yang-11prl}, which clearly indicates the breaking
down of the robust quantum spin Hall transport. In the case it is
understandable since there is no additional symmetry to protect the
QSHE.

The in-plane Zeeman field may couple two spin-components such that
we couldn't calculate the spin Chern numbers as we did in the previous
sections. It is still unclear whether or not there exists a hidden
symmetry similar to $S_{z}$ after proper transformation such that
we could define two ``spin''-dependent Chern numbers or the Zeeman
coupling is relatively negligible. This will be a subject we need
to study further. In the experiment by Du et al\cite{Du-13xxx} the
conductance is quantized while the Fermi level sweeps the whole band
gap by tuning the gate voltage. The quantum plateau is also robust
up to an 8T perpendicular field and a 12T in-plane magnetic field.
Any sub-gap of order of $10^{-3}$meV in the edge states (if exists)
should be measurable at low temperatures (tens mK) by sweeping the
Fermi level or the gate voltage. One possible explanation is that
the Rashba-like spin orbit coupling between $h_{+}$ and $h_{-}$
is negligible, and that the edge states are robust with respect to
the $S_{z}$ symmetry. In this case the Chern numbers in $h_{+}$
and $h_{-}$ are still well defined in the presence of a magnetic
field. Another possibility is that the strong disorder effect suppresses
the spin-orbit coupling. In the experiment the mobility gap and the
quantum plateau of conductance coexist, indicating the occurrence
of the Anderson localization for the bulk electrons. This is a clear
signature of topological Anderson insulator \cite{Li-09prl,Groth-09prl}.
Thus the disorder may stabilize the QSHE in the system. More studies
are expected alone this direction in the future.

In general, when time reversal symmetry is broken, the edge states
may open a sub-gap even if there is no bulk band gap closing, which
is different from topological quantum phase transition when the symmetry
is invariant. For instance the zero end modes in the the Su-Schrieffer-Heeger
model with the chirality symmetry move away from the zero energy to
the bulk band when a staggered on-site potential is introduced, which
breaks the chirality symmetry. In this case the bulk gap never closes
even when the end modes move into the bulk bands. This is a key to
understand the Thouless charge pumping in the Rice-Mele model\cite{Charge-pump}.

\section{Summary}

In short, the quantum spin Hall effect can persist up to a strong
magnetic field when $S_{z}$ is a good quantum number. In this case
the spin-dependent Chern number is well defined in each subspace,
and according to the bulk-edge correspondence\cite{Hatsugai-93prl},
the helical edge states are robust to an external field.

\section*{Acknowledgments}

This work was supported by the Research Grant Council of Hong Kong
under Grants No. 17304414.

\end{document}